# Improved visual detection of moving objects in astronomical images using color intensity projections with hue cycling

Keith S Cover
VU University Medical Center
Amsterdam, The Netherlands

## Abstract

While fully automated methods for detecting faint moving objects in astronomical images - such as Kuiper belt objects (KBOs) - are constantly improving, visual detection still has a role to play especially when the fixed background is cluttered with stars. Color intensity projections (CIPs) using hue cycling – which combines a sequence of greyscale images into a single color image – aids in the visual detection of moving objects by highlighting them using color in an intuitive way. To demonstrate the usefulness of CIPs in detecting faint moving objects a sequence of 16 images from the SuprimeCam camera of the Subaru telescope were combined into a CIPs image. As well has making even faint moving objects easier to visually detect against a cluttered background, CCD artefacts were also more easily recognisable. The new Hyper SuprimeCam for the Subaru telescope – which will allow many short exposure images to be acquired with little dead time between images – should provide ideal data for use with the CIPs algorithm. In addition, the current search for KBOs to be targeted by the New Horizon's spacecraft after its flyby of Pluto provides an excellent test case for the state of the art in faint moving object detection against a cluttered background.

**Introduction**

In astronomy the tried and true method of visually detecting moving objects against a fixed background is to toggle back and forth between two images. However, major improvements in astronomical instruments allows many images to be acquired over the same time interval as a single image with little loss of information. The Subaru Hyper SuprimeCam currently under construction is a prime example. Thus it may be possible to use many images when trying to detect a moving object.

While computer algorithms are commonly used to detect moving objects (Jewitt et al. 1998; Gladman et al. 1998; Allen et al. 2001; Petit et al. 2004) visual inspection is still often used in practice - in spite of it being both tedious and very time consuming - because the current algorithms still miss many objects especially against a cluttered background.

The ongoing hunt for Kuiper belt objects (KBOs) that the NASA's New Horizons spacecraft can target after flying by Pluto in July 2015 is an excellent example of current practices in detecting faint moving objects (Spencer et al. 2003, Guo et al. 2006). Current predictions are that only a few KBOs may be within range of a flyby - all with a brightness of 24 magnitude or less. Detecting these faint KBOs is made much more challenging by clutter introduced by the Milky Way galaxy, which is Pluto's current backdrop. However, KBO's motion at about 1 arcsecond per hour allows them to be differentiated from the fixed stars.

Up to date information on the New Horizon's KBO (NHKBO) search can be found at http://wikivm.boulder.swri.edu/mediawiki/index.php/New_Horizons_KBO_Search.

So far, the NHKBO search has acquired images with several telescopes including the 8.2m Subaru telescope located at the summit of Mauna Kea, Hawaii, USA. While the sky scanning protocol is constantly being reviewed and updated, fields have been visited up to 3 times per night with exposure times of about 30 minutes each. As SuprimeCam camera of the Subaru telescope has a pixel size of 0.2 arc seconds, a KBO will drift about 5 pixels per hour against the fixed background.

In addition to automated detection, visual inspection methods are being employed in the NHKBO search. Currently, the difference between two images is being inspected visually. However, stars with variable intensity far out number potential KBOs, greatly reducing the efficiency of the detection method.

Cover et al. (2007B) has proposed a method for combining any number of grey scale images into a single color image. In color intensity projections (CIPs), when a pixel has the same intensity in all the component images it has the same grey scale value in the combined image. However, if the intensity of the pixel changes over the images the combined pixel has a hue that encodes the time of the maximum pixel intensity. The algorithm presented here is a slightly simplified version of Cover et al. (2007) with the addition of hue cycling.

CIPs was originally developed for applications in medical imaging (Cover et al. 2006, 2007A) but the generality of the technique made it applicable to many applications. The algorithm presented herein is a slightly simplified version of Cover et al. (2006) with the addition of hue cycling.

The hue of color cycles over red-yellow-green-light blue-blue-purple-red. Cover et al. (2006, 2007A, 2007B) only used red through blue so that the hue would map to a unique time. However, there are situations, such detecting faint KBO's were, rapidly varying hue can be more useful than being able to map the hue to a unique time. Thus, hue is cycled over several periods in this paper.

When displaying a series of greyscale images as a single CIPs, an interesting question would be what number of visits for a fixed total time in a single night would yield the most likely detection of a faint KBO. For example is 3 visits of 30 minutes each spent visiting a single field more effective at detecting a faint KBO than 6 visits at 15 minutes each or 9 visits of 10 minutes each? While the shorter acquisition times per visit will decrease the signal to noise ratio of each image, it will also yield more hues in the combined image for each KBO. And more hues will improve the visual detection of the KBO against background clutter such as the Milky Way. Unfortunately, the precise simulations required to answer this question are beyond the scope of this paper.

Instead, to demonstrate the effectiveness of CIPs in detecting moving objects 16 images acquired consecutively by the Subaru telescope SuprimeCam are displayed as CIPs and slowly moving objects are detected.

**Method**

This study combined 16 images, which had been acquired consecutively with the same CCD of the SuprimeCam camera of the Subaru telescope into a single CIPs image. The images had acquisition times of 120s with 60s between exposures. The first exposure had image number SUPA00334604. Each CCD has 2080 x 4100 pixels with each pixel having a resolution of 0.202 x 0.202 arcseconds.

The Subaru telescope has an effective aperture of 8.2m (Miyazaki et al. 2002). The 16 images used in this study were downloaded from SMOKA (Subaru Mitaka Okayama Kiso Archive system). Development and operations of SMOKA are carried out by the Astronomy Data Center (ADC) of the National Astronomical Observatory of Japan (NAOJ) (Baba et al. 2002; Yamamoto et al. 2003; Enoki et al. 2004; Ideta et al. 2005; Yamada et al. 2009).

As the images were acquired the weather was degrading resulting in a bright background in the later images. To remove the background, each pixel in all images were high passed filtered. The filtering was implemented using ImageJ and subtracted from each pixel the median of all the pixels within 20 pixels of the pixel of interest.

A maximum intensity projection of the windowed 16 images was calculated to provide a comparison of the CIPs of the 16 images with conventional combination techniques.

The colored CIPs image was calculated from the unwindowed but high passed 16 greyscale images. The brightness, hue and saturation of the CIPs was calculated individually from the greyscale images on a pixel by pixel basis. The range of hue and saturation is typically zero to one. The CIPs image was calculated by the following simple equations

$$\begin{aligned}&Brightness\ = MaxIP,\\ &Saturation = (MaxIP\text{-}MinIP)\ /MaxIP \qquad\qquad (1)\\ &Hue\ =\ (image\ index\ of\ the\ maximum\ value\ for\ the\ pixel)/(images\ per\ hue\ cycle)\end{aligned}$$

Where MaxIP is the maximum value of the pixel over the greyscale images and MinIP is the minimum value of the pixel over the greyscale images. The image index is 0 for the first image, 1 for the second image, …, and 15 for the last image. The images per hue cycle is 6 for this example. For example, if a pixel had the maximum value in the first acquired image, its hue would be red.

After calculating the CIPs image but before displaying it the brightness was windowed by setting the brightness of any pixel that was above 200 counts to 200 counts.

To display the CIPs hue-saturation-brightness image it is usually necessary to convert it to an RGB image. While there are a variety of different ways to accomplish this conversion the algorithm used for this paper was the HSB to RGB conversion routine in the Java Programming Language.

Just as it is possible to scale the brightness and contrast of a greyscale image being displayed to enhance features of interest, it is also possible to scale both the hue and saturation of a CIPs image. The hue cycling discussed above allows the "contrast" of the hue to be adjusted.

The saturation can also be scaled by linearly transforming the saturation after it has been calculated with the above equation. Often for saturation, only the scale factor is needed. Scaling the saturation can give a better indication of how much the intensity of a pixel has varied over the component greyscale images. For a moving object on a black background the change in intensity is always 100% so the saturation is also 100%. However, if dealing with a variable star the saturation may be much smaller and there may be a desire to scale it up. Also, it is possible to take a log scale of the greyscale images before combining them into a CIPs to enhance the saturation due to particular changes in intensity.

**Results and Discussion**

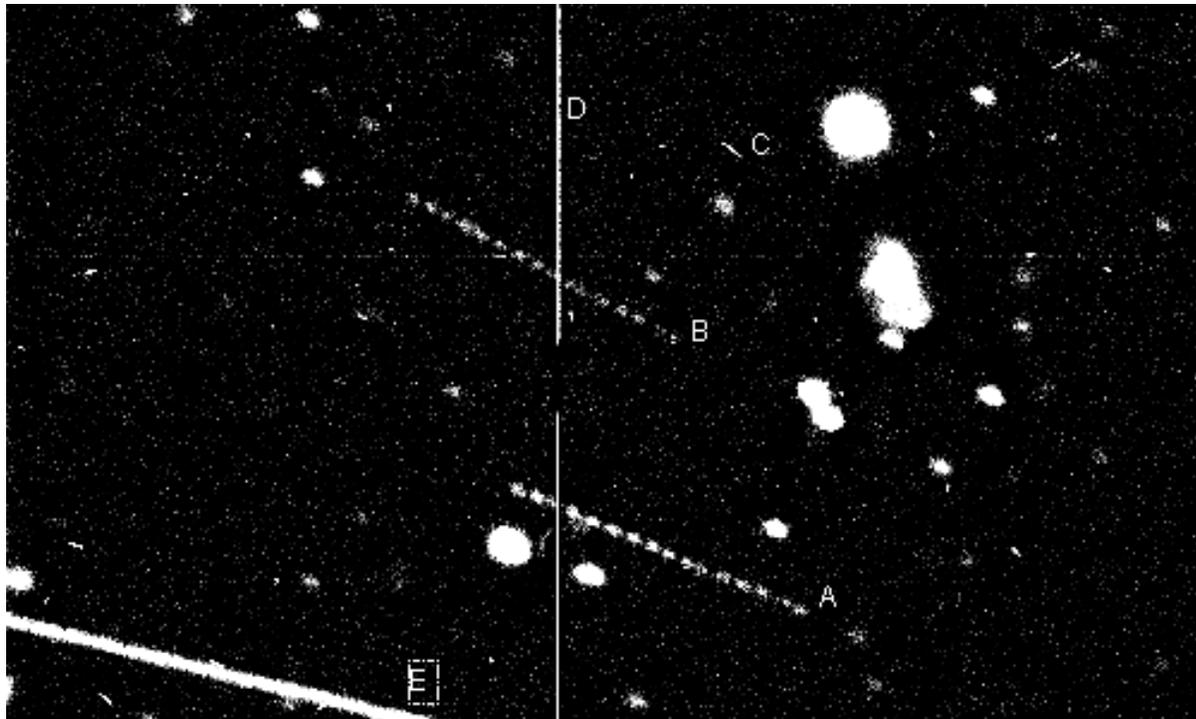

*Figure 1. Maximum intensity projection of a region of the 16 images. The letters A to E indicate some patterns that are of interest.*

Figure 1 shows the maximum intensity projection of the 16 images. Letters A through E might suggest moving objects. The vertical path denoted by letter D is clearly a CCD artefact. Patterns A and B are highly suggestive of objects moving over the 16 images including 16 separate points. The cause of the patterns indicated by letters C and E are less clear.

Figure 2 shows the CIPs of the 16 images for a larger region than Figure 1 but also includes it. Patterns A and B show the cyclic hue pattern of red-yellow-green-light blue-blue-purple-red of an object moving at a constant velocity. Each of the two patterns contains exactly 16 points, again an indication of an object moving at a constant velocity. Since the first image has the hue red both objects are travelling from the upper left to the lower right.

Patterns C and E are a single hue with C purple and E green. This almost certainly indicates that each are from a single image and likely some sort of image artifacts. Although the possibility that pattern E corresponds to a very fast moving object cannot be ruled out.

Pattern D is bright grey in color indicating the pattern is constant through all images. This consistency combined with its vertical orientation in the CCD confirm the obvious conclusion that pattern D is some sort of artefact of the CCD.

Patterns F, G and H in Figure 2 appear similar to pattern C in Figure 2 and are likely some sort of CCD artefact that only exists for a single image.

Several stars are obvious in Figure 2 and have a distinctive color pattern. The pattern are indicative of the apparent motion of the stars due to atmospheric distortions. The pattern is similar for adjacent stars since the atmospheric distortion is similar.

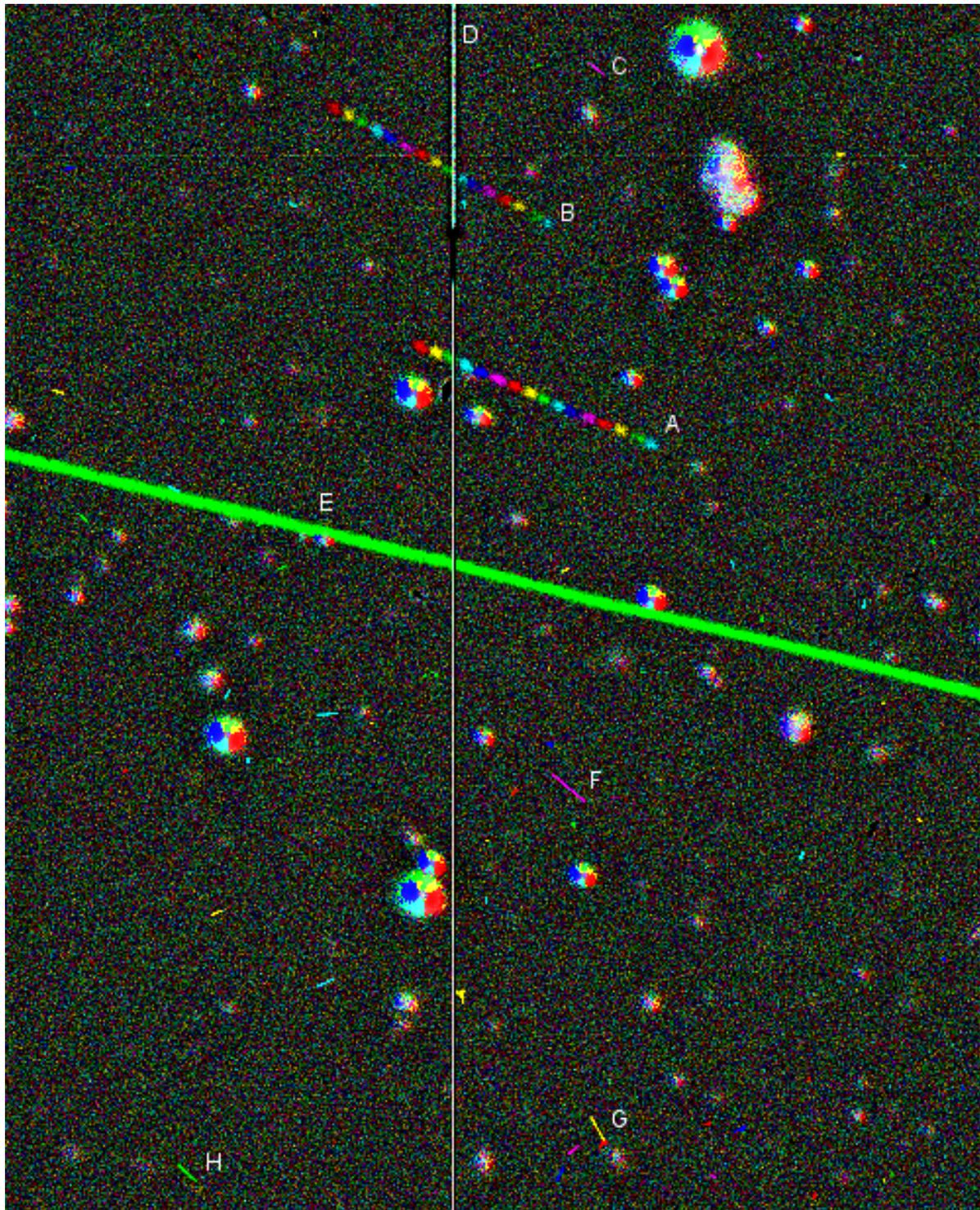

Figure 2. CIPs image including all the pixels of Figure 1 plus a region below Figure 1. The letters A to E correspond to the same patterns as Figure 1. Letters F to H indicate additional patterns.

One other feature worth mentioning in Figure 2 is the colorful noise. Since the noise is always changing from image to image it is expected it will be colorful in a CIPs image. Since the brightness of the pixels with noise, as with all other pixels, is the maximum of the same pixel in the 16 component images the noise appears larger in a CIPs image than the component image.

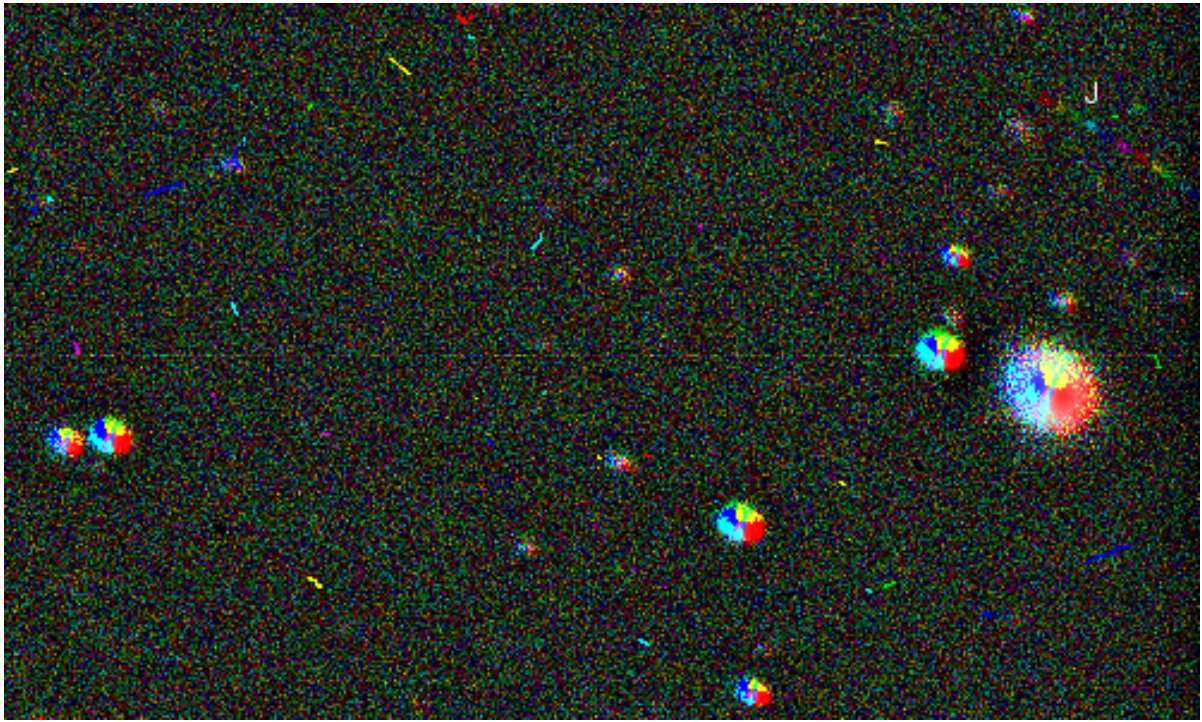

*Figure 3. Another region of the as CIPs image as shown in Figs 1 and 2. A faint moving object is indicated by the letter J.*

Figure 3 shows a different region of the same image displayed in Figures 1 and 2. Pattern J in the upper right corner shows a faint moving object near a group of stars. Even though the individual points in the pattern are faint the hue cycling pattern clearly stands out against the noise. The pattern would likely still standout, even against a crowded stellar background, even if it was fainter.

**Conclusions**

The CIPs algorithm is simple to implement and makes moving objects easier to visually distinguish from artefacts. However, to be effective CIPs likely requires at least 4 or more images against the same fixed background.

For detecting a faint moving object against a cluttered background with the total time fixed, the question still remains is how many images to acquire. Ether simulation or implanting moving objects in a sequence of real images will be required to determine this number.

**Acknowledgements**

This work was funded by the VU University Medical Center in Amsterdam.

**Disclosure:** The author's employer is pursuing a patent on color intensity projections covering applications in astronomy, medical imaging, and other fields. The author has a financial interest in the patent.

**REFERENCES**

Allen, R. L., Bernstein, G. M., & Malhotra, R. 2001, ApJ, 549, L241
Arnold, K., Gosling, J., & Holmes, D. 2005, Java Programming Language (4th ed.; Englewood Cliffs, NJ: Prentice-Hall)
Baba, H., Yasuda, N., Ichikawa, S., Yagi, M., Iwamoto, N., Takata, T., Horaguchi, T., Taga, M., Watanabe, M., Okumura, S., Ozawa, T., Yamamoto, N., and Hamabe, M. 2002, "Development of public science archive system of Subaru Telescope", RNAOJ., 6, 23-36
Biesecker, D. A., Lamy, P., St. Cyr, O. C., Llebaria, A., & Howard, R. A. 2002, Icarus, 157, 323
Buie, M. W., et al. 2003, Earth Moon Planets, 92, 113
Cover, K. S., Lagerwaard, F. J., & Senan, S. 2006, Int. J. Radiat. Oncol. Biol. Phys., 64, 954
Cover, K. S., Lagerwaard, F. J., van den Berg, R., Buis, D. R., & Slotman, B. J. 2007A, Neurosurgery, 60, 511
Cover, K. S., Lagerwaard, F. J., Senan, S. 2007B, Publications of the Astronomical Society of the Pacific, 119, 523–526
Enoki, M., Taga, M., Ozawa, T., Noda, S., Okumura, S., Yoshino, A., Furusho, R., Baba, H., Horaguchi, T., Takata, T., and Ichikawa, S. 2004, "Development of public science archive system of Subaru Telescope 3", RNAOJ., 7, 57-84
Gladman, B., Kavelaars, J. J., Nicholson, P. D., Loredo, T. J., & Burns, J. A. 1998, AJ, 116, 2042
Guo, Y.P., Farquhar, R.W., 2006 Acta Astronautica 58, 550
Ideta, M., Enoki, M., Ozawa, T., Yoshino, A., Nakata, F., Okumura, S., Yamamoto, N., Furusho, R., Yaji, K., Yamada, Y., Yagi, M., Horaguchi, T., Takata, T., and Ichikawa, S. 2005, "Development of public science archive system of Subaru Telescope 4", RNAOJ., 8, 59-84
Jewitt, D., Luu, J., & Trujillo, C. 1998, AJ, 115, 2125
Miyazaki, S., et al. 2002, Publications of the astronomical society of Japan, 6, 833
Morrill, J. S., et al. 2006, Sol. Phys., 233, 331
Petit, J.-M., Holman, M., Scholl, H., Kavelaars, J., & Gladman, B. 2004, MNRAS, 347, 471
Sekanina, Z. 2002, ApJ, 566, 577
Spencer, J., Buie, M., Young, L., Guo, Y.P., Stern, A., 2003, Earth Moon Planets, 92, 483-491
Veillet, C., et al. 2002, Nature, 416, 711
Yamada Y., Ozawa, T., Nishizawa A., Furusho R., Nishimura T., Enoki M., Yoshino A., Furusawa J., Takata T., and Ichikawa S. 2009, "Development of public science archive system of Subaru Telescope. 5", RNAOJ, 12, 53-78
Yamamoto, N., Noda, S., Taga, M., Ozawa, T., Horaguchi, T., Okumura, S., Furusho, R., Baba, H., Yagi, M., Yasuda, N., Takata, T., and Ichikawa, S. 2003, "Development of public science archive system of Subaru Telescope. 2", RNAOJ., 6, 79-100